\def\be{\begin{eqnarray}}
\def\en{\end{eqnarray}}

\def\B{{\cal B}}

\def\pr{{Phys. Rev.}~}
\def\prl{{ Phys. Rev. Lett.}~}
\def\pl{{ Phys. Lett.}~}
\def\np{{ Nucl. Phys.}~}

\documentclass{ws-ijmpa}

\begin{document}

\markboth{Hai-Yang Cheng} {PARITY-EVEN AND PARITY-ODD MESONS IN
COVARIANT LIGHT-FRONT APPROACH}

%
\catchline{}{}{}{}{}
%

\title{PARITY-EVEN AND PARITY-ODD MESONS IN COVARIANT LIGHT-FRONT APPROACH\\
}

\author{\footnotesize Hai-Yang Cheng}

\address{Institute of Physics, Academia Sinica\\
Taipei, Taiwan 115, ROC}

\maketitle


\begin{abstract}
Decay constants and form factors for parity-even ($s$-wave) and
parity-odd ($p$-wave) mesons are studied within a covariant
light-front approach. The three universal Isgur-Wise functions for
heavy-to-heavy meson transitions are obtained.

\end{abstract}

\section{Introduction}

Interest in even-parity charmed mesons has been revived by recent
discoveries~\cite{data} of two narrow resonances: $D_{s0}^*(2317)$
and $D_{s1}(2460)$, and two broad resonances, $D_0^*(2308)$ and
$D_1(2427)$. The unexpected and surprising disparity between
theory and experiment has sparked a flurry of many theory papers.

Before our work,\cite{CCH} the Isgur-Scora-Grinstein-Wise (ISGW)
quark model\cite{ISGW,ISGW2} is the only model that can provide a
systematical estimate of the transition of a ground-state $s$-wave
meson to a low-lying $p$-wave meson. However, this model is based
on the non-relativistic constituent quark picture.  Since the
final-state meson at the maximum recoil point $q^2=0$ or in
heavy-to-light transitions can be highly relativistic, it is thus
important to consider a relativistic approach. The covariant
light-front model elaborated in \cite{Jaus99} is suitable for this
purpose, but again it has been only applied to $s$- to $s$-wave
meson transitions. In \cite{CCH} we have extended the covariant LF
quark model to parity-even, $p$-wave mesons and studied their
decay constants, form factors and the corresponding Isgur-Wise
functions.

\section{Decay constants and form factors}

Consider the decay constants for mesons with the quark content
$q_1\bar q_2$ in the $^{2S+1} L_J= {}^1S_0$, $^3P_0$, $^3S_1$,
$^3P_1$, $^1P_1$ configurations. In the SU(N)-flavor limit
($m_1=m_2$) the decay constants $f_{S(^3P_0)}$ and $f_{^1P_1}$
should vanish.\cite{Suzuki}
In the heavy quark limit ($m_1\to\infty$), it is more convenient
to use the $L^j_J=P^{3/2}_2$, $P^{3/2}_1$, $P^{1/2}_1$ and
$P^{1/2}_0$ basis as the heavy quark spin $s_Q$ and the total
angular momentum of the light antiquark $j$ are separately good
quantum numbers.
Since decay constants should be identical within each multiplet,
$(S^{1/2}_0,\,S^{1/2}_1),\,(P^{1/2}_0,\,P^{1/2}_1),\,(P^{3/2}_1,\,P^{3/2}_2)$,
heavy quark symmetry (HQS) requires~\cite{IW89,HQfrules}
\begin{eqnarray} \label{eq:HQSf}
 f_V=f_P,\qquad
 f_{A^{1/2}}=f_S,\qquad
 f_{A^{3/2}}=0,
\end{eqnarray}
where we have denoted the $P^{1/2}_1$ and $P^{3/2}_1$ states by
$A^{1/2}$ and $A^{3/2}$, respectively. It is important to check if
the calculated decay constants satisfy the non-trivial
SU(N)-flavor and HQS relations. The numerical results are shown in
Table 1.

\begin{table}[h]
\tbl{Mesonic decay constants (in units of MeV) obtained. Those in
parentheses are taken as inputs.} {\begin{tabular}{|c|ccccc|}
\hline
 $^{2S+1} L_J$
          & $f_{u\bar d}$
          & $f_{s\bar u}$
          & $f_{c\bar u}$
          & $f_{c\bar s}$
          & $f_{b\bar u}$
          \\
\hline $^1S_0$
          & $(131)$
          & $(160)$
          & $(200)$
          & $(230)$
          & $(180)$
          \\
$^3P_0$
          & $0$
          & $22$
          & $86$
          & $71$
          & $112$
          \\
$^3S_1$
          & $(216)$
          & $(210)$
          & $(220)$
          & $(230)$
          & $(180)$
          \\
$^3P_1$
          & $(-203)$
          & $-186$
          & $-127$
          & $-121$
          & $-123$
          \\
$^1P_1$
          & $0$
          & $11$
          & $45$
          & $38$
          & $68$
          \\
\hline $P^{1/2}_1$
          & --
          & --
          & $130$
          & $122$
          & $140$
          \\
$P^{3/2}_1$
          & --
          & --
          & $-36$
          & $-38$
          & $-15$
          \\
\hline
\end{tabular}}
\end{table}

From Table 1 we see that the decay constants of light scalar
resonances are suppressed relative to that of the pseudoscalar
mesons, while the suppression becomes less effective for heavy
scalar mesons. Our result of $f_{D_{s0}^*}=71$ MeV is supported by
the measurements of the $B\to D^{(*)}\bar D_{s0}^*$
decays.\cite{data}

Form factors for heavy-to-heavy and heavy-to-light transitions
have been computed in the covariant light-front approach. The
details are shown in \cite{CCH}. Our results for form factors in
$B\to D,D^*,D^{**}$ ($D^{**}$ denoting generic $p$-wave charmed
mesons) transitions agree with those in the ISGW2
model.\cite{ISGW2} Relativistic effects are mild in $B\to D$
transition, but they could be more prominent in heavy-to-light
transitions, especially at maximum recoil ($q^2=0$). For example,
we obtain $V_0^{Ba_1}(0)=0.13$,\cite{CCH} while ISGW2 gives 1.01.
If $a_1(1260)$ behaves as the scalar partner of the $\rho$ meson,
it is expected that $V_0^{Ba_1}\sim A_0^{B\rho}\sim O(0.3)$. The
predicted decay rates for $\overline B\to D^{**}\pi$ and
$\overline D_s^{**} D^{(*)}$ obtained in the CLF model agree with
experiment.\cite{CCH}

It is worth mentioning that the ratio
 $R=\B(B^-\to D_2^{*0}\pi^-)/\B(B^-\to D_1^0\pi^-)$
is measured to be $0.80\pm0.07\pm0.16$ by BaBar \cite{BaBarD} and
$0.77\pm0.15$ by Belle.\cite{BelleD}  The early prediction by
Neubert \cite{Neubert} yields a value of 0.35, while
soft-collinear effective theory predicts $R=1$.\cite{Mantry} Our
prediction of $R=0.91$ in the covariant light-front model is in
accordance with the data.

\section{Heavy quark limit and Isgur-Wise functions}

In the heavy quark limit, heavy quark symmetry\cite{IW89} provides
model-independent constraints on the decay constants and form
factors. For example, pseudoscalar and vector mesons would have
the same decay constants and all the heavy-to-heavy mesonic decay
form factors are reduced to some universal Isgur-Wise functions.
Therefore, it is important to study the heavy quark limit behavior
of these physical quantities to check the consistency of
calculations.

It is well known that the $s$-wave to $s$-wave meson transition in
the heavy quark limit is governed by a single universal IW
function $\xi(\omega)$.\cite{IW89} Likewise, there exist two
universal functions $\tau_{1/2}(\omega)$ and $\tau_{3/2}(\omega)$
describing ground-state $s$-wave to $p$-wave
transitions.\cite{IW91} The calculated IW functions are shown in
Fig. 1. It is found that at zero recoil $\omega=1$, $\xi(1)=1$,
$\tau_{1/2}(1)=0.61$, $\tau_{3/2}(1)=0.31$ and $\rho^2=1.22$ for
the slope parameter of $\xi(\omega)$. Our results for $\tau_{1/2}$
and $\tau_{3/2}$ agree well with the recent lattice results
\cite{Becirevic} $\tau_{1/2}(1)=0.38\pm0.05$ and
$\tau_{3/2}(1)=0.58\pm0.08$. The Bjorken \cite{Bjorken} and
Uraltsev \cite{Uraltsev} sum rules for the Isgur-Wise functions
are found to be fairly satisfied.

\begin{figure}[t]
 \centerline{
 \includegraphics[angle=-90,width=5.5cm]{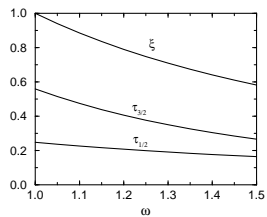}}
\caption{The Isgur-Wise functions $\xi$, $\tau_{1/2}$ and
$\tau_{3/2}$ as a function of $\omega$.}
\end{figure}

\section{Acknowledgments}
I am grateful to Chun-Khiang Chua and Chien-Wen Hwang for fruitful
collaboration.

\end{document}